\documentstyle[preprint,aps]{revtex}
\begin{document}
\preprint{IUCM96-007}
\draft
%\twocolumn[
%\widetext

\title{Long-lived charged multiple-exciton complexes in strong magnetic
fields}
\author{J. J. Palacios, D. Yoshioka\cite{byline}, and  A. H. MacDonald}
\address{Department of Physics, Indiana University, Bloomington, IN 47405,
USA.}

\date{\today}
\maketitle

\widetext
\begin{abstract}
\leftskip 2cm
\rightskip 2cm
We consider the charged exciton complexes of an ideal two-dimensional
electron-hole system in the limit of strong magnetic fields. A series of
charged multiple-exciton states is identified and variational and
finite-size exact diagonalization calculations are used to estimate
their binding energies.  We find that, because
of a hidden symmetry, bound states of excitons and an additional electron
cannot be created by direct optical absorption and, once created, have an
infinite optical recombination lifetime.  We also estimate the optical
recombination rates when electron and hole layers are displaced and the
hidden symmetry is violated.

\end{abstract}

%]
%\narrowtext

Two-electron atoms are among the simplest systems in which electronic
exchange and correlation play an important role.  Studies of these
systems\cite{bethe} have played a vital role in the development of
practical techniques for accurate calculations in many-electron atoms and
molecules.  Among two-electron atomic systems the $H^{-}$ ion, which is
barely bound, is the most difficult to describe accurately.  The
semiconductor analogs\cite{lampert} of the $H^{-}$ ion are the charged
exciton states, $X^{-}$ and $X^{+}$, for which, respectively,
two electrons are bound to a
single hole and two holes are bound to a single electron.
Recently there has been considerable experimental
interest\cite{french,barjoseph,pepper,dminus} in the charged exciton
states of two-dimensional (2D) electron systems, in part because the
reduced dimensionality leads to relatively larger binding energies.  At
zero magnetic field, the charged exciton has a single spin-singlet bound
state and the binding energy seen in 2D systems in recent experiments is
in reasonable agreement with theory.\cite{xminustheory} Experiments have
also demonstrated that for 2D charged excitons an additional spin-triplet
bound state becomes stable in sufficiently strong external magnetic
fields.  In this Letter we address the strong magnetic field limit for 2D
charged multiple-exciton complexes.\cite{pawel} We find that in addition
to the spin-triplet charged single-exciton states there exist a series of
charged bound multiple-exciton complexes.  We also find that, because of a
hidden symmetry in the Hamiltonian of this system, these states have
an infinite optical recombination time for an ideal system.  We propose
that this unanticipated anomaly should lead to observable effects in
time-resolved photoluminescence experiments.

In the strong magnetic field limit we consider, all particles in the
charged exciton complex are confined to their lowest Landau levels and
have their spins aligned with the magnetic field.  Our basic conclusions
follow from an exact mapping\cite{mapping,oldmapping} between spin
polarized particle-hole systems and spin-$1/2$ electron systems, which
holds in this limit.  Our analysis exploits recent advances\cite{skyrmions} in
understanding the elementary charged excitations of the incompressible
ground state in spin-$1/2$ electron systems at Landau level filling factor
$\nu \equiv N/ N_{\phi} =1 $. [Here $N_{\phi} = A  / (2 \pi \ell^{2}) =A
B/ \Phi_{0} $ is the Landau level degeneracy, $\ell$ is the magnetic
length and $\Phi_{0}$ is the electronic magnetic flux quantum.] In the
mapping the hole Landau level is associated with the minority
($\downarrow$) spin Landau level and the occupied states in the electron
Landau level are associated with empty states in the majority spin
($\uparrow$) Landau level\cite{caveat}:
\begin{eqnarray}
 N_{h} &=& N_{\downarrow} \nonumber\\
 N_{e} &=& N_{\phi} - N_{\uparrow}.
\label{eq:nhne}
\end{eqnarray}
 For example, charge neutral states ($N_{e}= N_{h}$) of electron-hole
systems correspond to $\nu =1$ ($ N = N_{\uparrow} + N_{\downarrow} =
N_{\phi} $ ) states of the spin-$1/2$ system.  Generally the total charge
($Q \equiv N_{e} - N_{h}$) and total particle-number ($L \equiv N_{e} + N_{h}$)
of the electron-hole system are given by $Q = N_{\phi} - N$ and $L = Q + N
- 2 S_{z}$ where $2 S_{z} = N_{\uparrow} - N_{\downarrow}$.  $X^{-}$
states have $Q=1$ and $L=3$ so they correspond to spin-$1/2$ states with
one particle removed from a full Landau level and $S_{z} = N/2 - 1$. The
eigenstates of electron-hole and spin-$1/2$ systems have a one-to-one
correspondence under this mapping\cite{mapping} and corresponding
eigenenergies differ by a known constant:
\begin{equation}
 E_{eh} = \tilde{E}_{1/2} - N_{e} I
\label{eq:energymapping}
\end{equation}
 where $I$ is the binding energy of an isolated exciton in this limit.
[For an ideal 2D system $I = (\pi/2)^{1/2} (e^{2}/\ell)$.]  Here all
energies are measured with respect to the corresponding non-interacting
electron values which increase with field in proportion to the quantized
kinetic energies of the Landau levels and
$\tilde{E}_{1/2}$ is the energy of the spin-$1/2$ system measured with
respect to the energy of the fully spin-polarized $\nu = 1$ state.

Recently, progress has been made in understanding the elementary charged
excitations of the $\nu =1$ ($N = N_{\phi}$) ground state for spin-$1/2$
particles.  This ground state has total spin quantum number $S = N/2$ and is
therefore spin-aligned by an arbitrarily weak magnetic field. Its charged
excitations have the unusual property, first noticed in numerical exact
diagonalization calculations\cite{rezayi} and dramatically evident in
recent experiments\cite{barrett}, that they can carry a large spin. It can
be shown\cite{hcm} that, for $N=N_{\phi}-1$, a single low-energy
spin-multiplet with orbital degeneracy $\approx N_{\phi}$ and energy
$\tilde{E}_{1/2}= \epsilon_{K}$ occurs with $S = N/2 -K$ for each $K = 0,
1, \cdots $.
$\epsilon_{K=0}=I$ and for large $K$ these elementary charged excitations
of the $\nu =1$ state can be identified\cite{skyrmions,hcm,jacob}
with the topologically charged Skyrmion spin-textures of the underlying
ferromagnetic $\nu =1$ ground state\cite{sondhi}.
In the $K \to \infty$ limit $
\epsilon_{K} \to 3 I /4 $. Numerical Hartree-Fock\cite{hf} and exact
diagonalization\cite{he} calculations indicate that $\epsilon_{K}$
decreases monotonically with $K$ between these limits.  Using the mapping
, since $S_{z} =N/2 -1$ states occur in both $S=N/2$ and
$S=N/2-1$ multiplets, it follows that
 for $N_{e} =2$ and $N_{h} =1$ there are two
low-lying states with energies $ E_{eh} = \epsilon_{K=0}-2I = -I $ and
$E_{eh} = \epsilon_{K=1} - 2 I $. The first of these states corresponds to
a single exciton and an unbound electron ($X+e$) while the second has
lower energy and  corresponds to a single exciton bound to an electron
($X^{-}_{1}$) with binding energy $\epsilon_{K=0} - \epsilon_{K=1}$. {\it
There are no other bound states between an electron and a single
exciton.}\cite{singletcaveat} (This lone
$X^{-}_{1}$ bound state in the strong magnetic field limit contrasts with
the solitary singlet and three triplet\cite{dminustheory} bound $D^{-}$
states for two 2D electrons bound to an external charge.)
The same analysis can be carried out for
larger values of $K$ and has unexpected implications: $K$ excitons and an
additional electron form a bound $X_{K}^{-}$ complex with energy
$(\epsilon_{K}-\epsilon_{K=0})-K I $.  The ionization energy of this
complex is
$\epsilon_{K=0} - \epsilon_{K}$ and dissociation energy is
$\epsilon_{K} - \epsilon_{K-1}$ for the reaction
$X_{K}^{-} \to X_{K-1}^{-} + X $.  Note that without the excess charge
there are no bound multiple-exciton complexes in the strong magnetic
field limit. (All binding energies are independent of
both electron and hole masses in the strong magnetic field limit.)

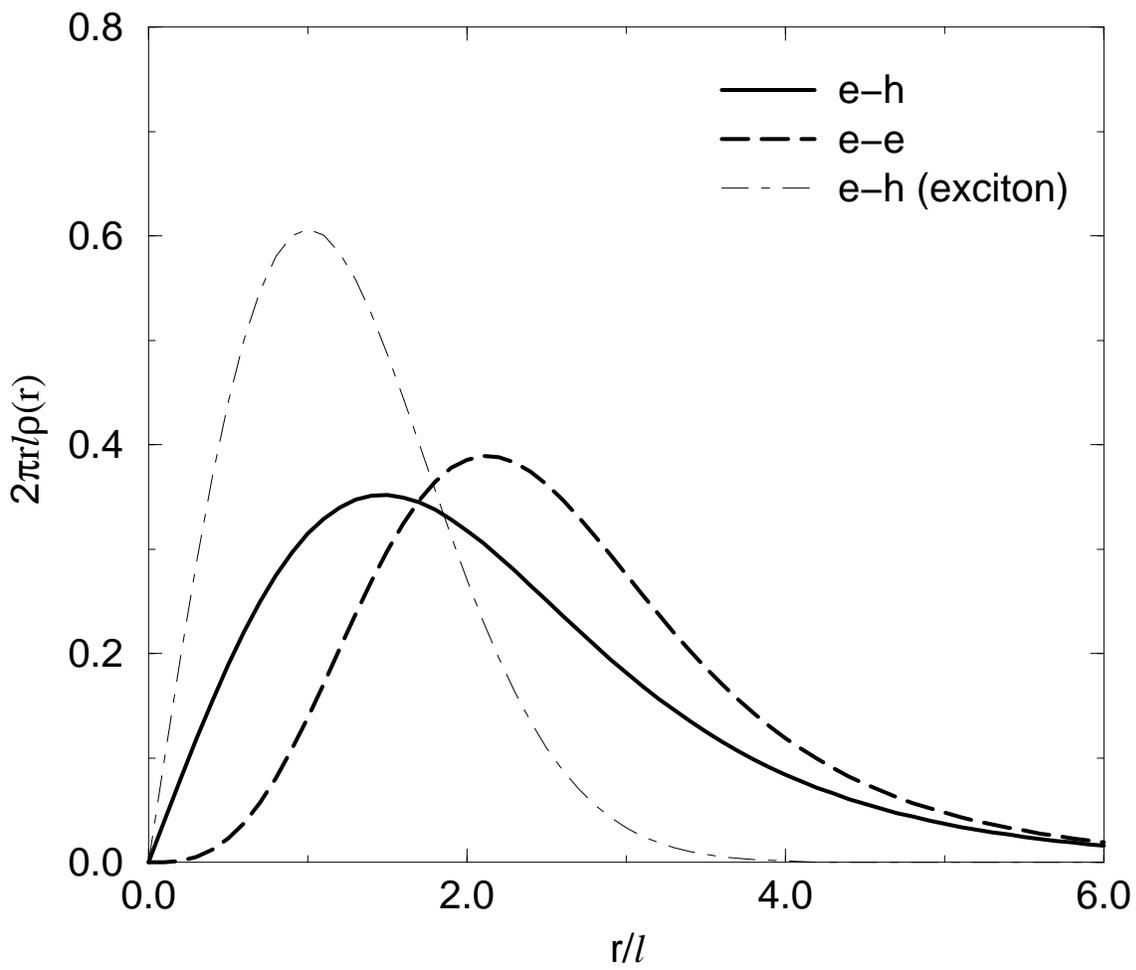
\begin{figure}
\caption{Electron-electron and electron-hole correlations in
$X^{-}_1$ state and an isolated exciton.  These graphs show the hole
density and the density of the other electron with one electron fixed at
the origin.  Note that the hole density at the origin is finite while the
other electron density vanishes.}
\label{fig:one}
\end{figure}

To estimate binding energies and optical matrix elements we perform
microscopic calculations using the symmetric gauge which has 
single-particle states with definite angular momentum in electron and
hole Landau levels.  The wavefunction which
describes the state $X+e$ (an exciton and an unbound electron) is
given in the corresponding occupation number representation by\cite{mapping} 
\begin{equation}
 |\Psi_{X+e}\rangle = \sum_{m=1}^{N_{\phi}} e_{m}^{\dagger}
h_{m}^{\dagger}  e_{0}^{\dagger} |0\rangle,
\label{eq:varwfxe}
\end{equation}
where $e^{\dagger}_{m}$ creates an electron with angular momentum
$-m$ and $h_{m}^{\dagger}$ creates a hole with angular momentum $m$.
Estimates of the binding energies of the $X_K^{-}$ complexes
for small $K$ can be obtained by using 
the following variational wavefunctions: 
\begin{equation}
|\Psi_{X^{-}_{K}}\rangle=\sum^{N_{\phi}}_{m_{K}>\cdots>m_{1}=0}
[\prod_{k=1}^{K}
\frac{a_{m_{k}}}{\sqrt{(m_{k}+1)}}
e^{\dagger}_{m_{k}+1}h^{\dagger}_{m_{k}}] e^{\dagger}_{0}|0\rangle.
\label{eq:varwfxm}
\end{equation} 
This form for the variational wavefunction is motivated by the fact that
for $a_{m}$ independent of $m$ it becomes exact\cite{hcm} in the case of
delta function repulsive interactions between the electrons and
attractive interactions between the electrons and the hole.  The
wavefunctions with constant $a_m$ also become exact in the large $K$
limit where they correspond in the spin language to the classical field
theory Skyrmions\cite{hcm,jacob}.  For the physically relevant case of
Coulombic electron-electron and electron-hole interactions we let $a_{m}
\propto \exp ( - \lambda m)$ where $\lambda$ is a variational parameter.
This form of the wavefunctions is motivated by our expectation that
longer-range interactions would favor more compact bound states and by
comparison with small system exact diagonalization
calculations.\cite{am} Table I shows the binding energies for the
simplest  $X^{-}_{K}$ complexes calculated from the wavefunctions
(\ref{eq:varwfxm}) by optimizing the variational parameter $\lambda$. As
expected, the exponential factor becomes irrelevant for $K\rightarrow
\infty$ where the classical field theory energy becomes exact.  The
validity of this wavefunction for $K=1$ has been checked by comparing
with results obtained by exact diagonalization of the Hamiltonian. It is
possible to diagonalize the Hamiltonian of a system sufficiently large
($N_{\phi} \le 80$) to make finite-size corrections negligible.  The
exact binding energy obtained in this way, $0.0545 (e^{2}/\ell)$, is
quite close to the variational one, $0.0529 (e^{2}/\ell)$.  Moreover,
the overlap of the variational wavefunction (\ref{eq:varwfxm}) for $K=1$
with the exact one is about 99\%. The estimate for $K=1$ is also in
qualitative agreement with existing experiments. Electron-electron and
electron-hole correlation functions for  the $X^{-}_{1}$ state are
illustrated in Fig. ~\ref{fig:one} and compared with the electron-hole
correlation function of an isolated bound exciton.  In this figure we
plot the probability of finding one electron or  the hole at radius $r$
when the other electron is at the origin. We see from this figure that
the quantum mechanical sharing of the hole among the two electrons makes
binding possible.  No binding occurs if the electron and hole are
treated as  classical particles.  The spin alignment of the two
electrons guarantees that the electrons do not have a large probability
of being close together;  in this strong field limit,  there is no bound
$X^{-}$ when the electrons form a  singlet even if the Zeeman energy is
discounted. Note that since the $X^{-}_{1}$ is a charged particle its
states occur in manifolds with degeneracy $ \sim N_{\phi}$ like the
Landau level manifolds for an isolated electron.

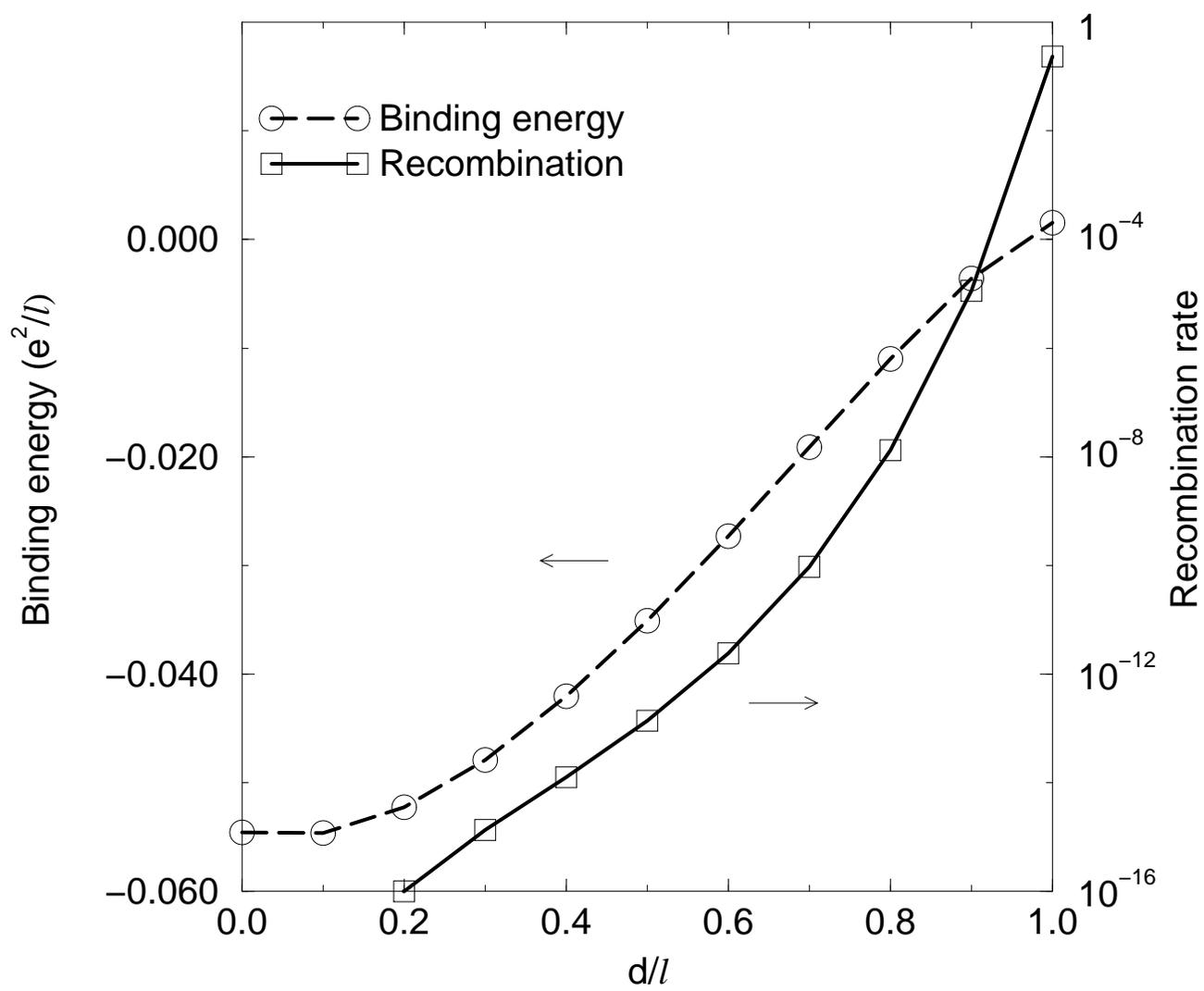
\begin{figure}
\caption{Normalized optical recombination rates and binding energy as a
function of electron-hole layer separation.  The binding energy is in
units of $(e^{2}/\ell)$.  These results were obtained from exact
diagonalization calculations for $N_{\phi}=50 $. The recombination rates at
small $d/\ell$ for this value of
$N_{\phi}$ are still decreasing with increasing $N_{\phi}$.}
\label{fig:two}
\end{figure}

The mapping between spin-$1/2$ and electron-hole systems has interesting
implications for the optical recombination matrix elements of
$X^{-}_{K}$ states. Let us focus again on the simplest $X^{-}_{1}$
state. As mentioned above, the eigenstates of the spin-$1/2$ system
occur in spin-multiplets because of the spin-rotational invariance of
the Hamiltonian and the unbound $X+e$ and bound $X^{-}_{1}$ states
correspond, respectively, to the $S_{z} = N/2 -1$ members of the $S=
N/2$ and $S = N/2 -1$ multiplets. The optical recombination operator, $R
= \sum_{m} e_{m} h_{m}$, maps\cite{mapping} to the total spin-raising
operator of the spin-$1/2$ system and therefore {\em annihilates} the
bound $X^{-}_{1}$ state.  In this ideal model the optical recombination
time of the $X^{-}_{1}$ state is infinite\cite{orthog}.  In real systems
there are a number of effects which will make the optical recombination
time finite, including Landau level mixing effects at finite magnetic
field strengths which depend in detail on the complicated valence band
electronic structure in a quantum well. Even at infinite field the
hidden symmetry is violated when electron-hole interactions differ from
electron-electron interactions by more than a change of sign, i.e.,
when the envelope functions for electron and hole Landau layers are not
identical.  This is often the case in quantum-well electron-hole
layers\cite{heiman}, for example when electrons and holes are
intentionally displaced by an electrostatic potential to inhibit optical
recombination.  In Fig. ~\ref{fig:two} 
we show results for the optical recombination
rate of $X^{-}_{1}$ states obtained from exact diagonalization
calculations for the case of narrow electron and hole layers separated
by a distance $d$.  The $X^{-}_{1}$ recombination rate is normalized to
the recombination rate of an isolated exciton. In this figure the
binding energy of $X^{-}_{1}$ is also shown. The $X^{-}_{1}$ state is
bound only for $d < \ell$. Results are shown for $N_\phi=50$.  The
recombination  rate at small $d/\ell$ is small and still decreasing with
system size.  In realistic experimental situations, the population of
excitons and charged excitons present in the system at strong magnetic
fields will depend in part on kinetic effects not discussed here.  Both
excitons and charged excitons will tend to be localized; the finite-size
effects we find in our calculations indicate that the actual
recombination rate  of a charged exciton will be sensitive to the
disorder potential it experiences.  Nevertheless, it follows from our
work that a dramatic difference between neutral and charged exciton
recombination times should occur and be observable in time-resolved
photoluminescence experiments.

This work has been supported by the National Science Foundation under
grant DMR-9416902.  J.J.P. acknowledges support from NATO postdoctoral
research fellowship, thanks L. Quiroga and C. Tejedor for bringing
his attention to Ref.\onlinecite{pawel}, and  thanks L. Mart\'{\i}n-Moreno
and J. H. Oaknin. for helpful discussions.
AHM acknowledges helpful conversations with Pawel Hawrylak and Don Heiman.

\begin{table}
\caption{Binding energies for the simplest $X^{-}_{K}$ exciton complexes.
The variational values have been obtained by optimizing the parameter
$\lambda$.  The binding energy for $K \to \infty$ is related  via the
mapping to the energy of Skyrmion states in quantum Hall ferromagnets.
All energies are in units of $(e^{2}/\ell)$.}
\begin{tabular}{|c|cccc|}
$ K $ & $\lambda$ & Variational binding energy\\
\hline
1&  0.071  & 0.0529\\
2&  0.045  & 0.0828\\
3&  0.034 & 0.1018\\
4& $\approx$ 0.03  &  $\approx$ 0.117\\
$\infty$ & 0  & 0.313\\
\end{tabular}
\end{table}

\begin{references}
\bibitem{caveat} In the version of the mapping in 
Ref.~\onlinecite{mapping} the particle-hole
transformation was performed for minority spins.
\bibitem{am} The classical field theory analysis in Ref.~\onlinecite{sondhi}
suggests that $a_m \propto \exp(-\lambda\sqrt{m})$ for 
large $m$ and we do find evidence for this in numerical exact diagonalization
calculations.  However, the present choice gives a better description
at shorter distances which dominate energy contributions.
\bibitem{orthog} The variational wave function given in
Eq.~(\ref{eq:varwfxm}) is not an exact eigenstate of  the mapped total
spin operator and so does not share this property.
\end{references}
\end{document}